\documentclass[12pt]{article}   
\usepackage[numbers]{natbib}
\usepackage{a4}
\usepackage{amsmath}
\usepackage{amssymb}
\usepackage{latexsym}
\usepackage{amssymb}
\usepackage{natbib}
\def \TP{{\mathrm{P}}}

\def \pd{\partial}

\def \e{{\mathrm{e}}}

\def \Bx{{\boldsymbol{x}}}

\begin{document}
\title{{\bf Non-singular dislocation loops in gradient elasticity   
}}
\author{
Markus Lazar~$^\text{a,b,}$\footnote{
{\it E-mail address:} lazar@fkp.tu-darmstadt.de (M.~Lazar).} 
\\ \\
${}^\text{a}$ 
        Heisenberg Research Group,\\
        Department of Physics,\\
        Darmstadt University of Technology,\\
        Hochschulstr. 6,\\      
        D-64289 Darmstadt, Germany\\
${}^\text{b}$ 
Department of Physics,\\
Michigan Technological University,\\
Houghton, MI 49931, USA
}

\date{\today}    
\maketitle


\begin{abstract}
 
Using gradient elasticity, 
we give in this Letter the non-singular fields produced 
by arbitrary dislocation loops in isotropic media.
We present the `modified' Mura, Peach-Koehler and Burgers formulae in 
the framework of gradient elasticity theory. 
\\

\noindent
{\bf Keywords:} dislocation loops; gradient elasticity.\\
\end{abstract}

We investigate the general theory of curved dislocations in isotropic media.
This topic is important in nano-mechanics of dislocations 
and can be found in any standard book on dislocation theory 
using the classical theory of elastostatics~\citep{Lardner,HL,Teodosiu,Mura,Li}.
The key-formulae in the `classical theory' of a closed dislocation loop $L$ 
are the 
Mura formula for the elastic distortion tensor
\begin{align}
\label{B-grad-0}
\beta^0_{ij}(\Bx)&=-\frac{b_k}{8\pi}\oint_L
\Big[\big(\epsilon_{jkl}\delta_{ir}-\epsilon_{rkl}\delta_{ij}
+\epsilon_{rij}\delta_{kl}\big)\pd_l \Delta
+\frac{1}{1-\nu} \,\epsilon_{rkl}\pd_l\pd_i\pd_j \Big] R\, 
d L'_r\,,
\end{align}
the Peach-Koehler formula for the stress tensor
\begin{align}
\label{T-grad-0}
\sigma^0_{ij}(\Bx)&=-\frac{\mu b_k}{8\pi}\oint_L
\Big[\big(\epsilon_{jkl}\delta_{ir}
+\epsilon_{ikl}\delta_{jr}\big)\pd_l \Delta
+\frac{2}{1-\nu}\, \epsilon_{rkl}\big(\pd_i\pd_j-\delta_{ij}\Delta\big)\pd_l
\Big] R\, 
d L'_r
\end{align}
and the Burgers formula for the displacement vector  
\begin{align}
\label{u-Burger-grad-0}
u^0_i(\Bx) = \frac{b_i}{8\pi}\, \int_S \Delta\pd_j R\, d S'_j
+\frac{b_l\epsilon_{rlj}}{8\pi}\, \oint_L
\bigg\{\delta_{ij} \Delta -\frac{1}{1-\nu}\, \pd_i \pd_j   
\bigg\}\, R\,  d L'_r\, ,
\end{align}    
where $R = |\Bx-\Bx'|$, $\mu$ is the shear modulus, $\nu$ is Poisson's ratio, 
$b_i$ is the Burgers vector of the dislocation line element $d L'_r$ at
$\Bx'$ and $d S'_j$ is the dislocation loop area.
The surface $S$ is the dislocation surface which is a cap of the 
dislocation line $L$.
These equations give the elastic fields and the displacement produced by a 
dislocation loop in isotropic media. 
They are valid only in the far-field due to singularities  
based on the unphysical description of the dislocation core as singular
delta functions. The purpose of this Letter is to 
give straightforward expressions for the dislocation fields 
which are free from singularities

A straightforward framework to obtain non-singular fields of dislocations
is the so-called theory of gradient elasticity. 
Particularly in a simplified, robust and often-used strain gradient elasticity
called gradient elasticity of Helmholtz type
the strain energy density has the form~\citep{LM05,GM}
\begin{align}
\label{W}
W=\frac{1}{2}\, C_{ijkl}\beta_{ij}\beta_{kl}
+\frac{1}{2}\, \ell^2 C_{ijkl}\pd_m \beta_{ij} \pd_m \beta_{kl}\, ,
\end{align}
where $C_{ijkl}$ is the tensor of elastic moduli,
$\beta_{ij}=\pd_j u_i -\beta^\TP_{ij}$ is the elastic distortion tensor,
$u_i$ and $\beta^\TP_{ij}$ denote the displacement vector and the plastic
distortion tensor respectively and $\ell$ is the material length scale 
parameter of gradient elasticity. For dislocations, $\ell$ is related to
the dislocation core radius.
Gradient elasticity is a continuum model of dislocations with 
core spreading.
Non-singular fields of straight dislocations 
were obtained in the framework of gradient elasticity of Helmholtz type
by~\citet{GA99,LM05,LM06} and \citet{Gutkin06}.
Surprisingly, not a single work has been done up to now in the direction of 
non-singular dislocation loops using strain gradient elasticity theory.
The reason lies probably in the expected mathematical complexity of the problem.

For an isotropic material
the tensor of elastic moduli reduces to
\begin{align}
\label{C}
C_{ijkl}=\mu\Big(
\delta_{ik}\delta_{jl}+\delta_{il}\delta_{jk}
+\frac{2\nu}{1-2\nu}\, \delta_{ij}\delta_{kl}
\Big)\,.
\end{align}
As shown by~\citet{LM05,LM06} the following governing equations 
for the displacement vector and the elastic distortion tensor can be derived
from the framework of gradient elasticity of Helmholtz type
\begin{align}
\label{u-H}
&L\, u_i =u_i^0\,,\\
\label{B-H}
&L\, \beta_{ij} =\beta_{ij}^0\,,
\end{align}
where $L=1-\ell^2\Delta$ is the Helmholtz operator. 
The singular fields $u_i^0$ and $\beta_{ij}^0$ are the sources
in the inhomogeneous Helmholtz equations (\ref{u-H}) and (\ref{B-H}).
The Helmholtz equations (\ref{u-H}) and (\ref{B-H}) can be further reduced to
inhomogeneous Helmholtz-Navier equations
\begin{align}
\label{u-L}
&L\, L_{ik} u_k =C_{ijkl}\pd_j \beta^{\TP,0}_{kl}\,,\\
\label{B-L}
&L\, L_{ik}\beta_{km} =-C_{ijkl}\epsilon_{mlr}\pd_j \alpha_{kr}^0\,,
\end{align}
where $L_{ik}=C_{ijkl}\pd_j\pd_l$ is the differential operator of the Navier 
equation.  In Eqs.~(\ref{u-L}) and (\ref{B-L}) the source terms 
are now the plastic distortion $\beta^{\TP,0}_{kl}$ and the 
dislocation density $\alpha_{kr}^0$ known from classical elasticity.
For a dislocation loop, they read~\citep{deWit}
\begin{align}
\label{A0}
\alpha^0_{ij}&=b_i\, \delta_j(L)=b_i \oint_L \delta(\Bx-\Bx')\, d L'_j\,,\\
\label{B0}
\beta^{\TP, 0}_{ij}&=-b_i\, \delta_j(S)=-b_i \int_S \delta(\Bx-\Bx')\, d S'_j\,.
\end{align}

The corresponding three-dimensional Green tensor 
of the Helmholtz-Navier equation is defined by
\begin{align}
&L\, L_{ik}\, G_{kj} =-\delta_{ij}\, \delta(\Bx-\Bx')
\end{align}
and is calculated as
\begin{align}
\label{G}
G_{ij}(R)=\frac{1}{16\pi\mu(1-\nu)}\, \Big[2(1-\nu)\delta_{ij}\Delta-
\pd_i\pd_j\Big] A(R)\,.
\end{align}
Hence
\begin{align}
\label{A}
A(R)=R+\frac{2\ell^2}{R}\,
\Big(1-\e^{-R/\ell}\Big)\, .
\end{align} 
In the limit $\ell\rightarrow 0$, the three-dimensional Green tensor
of classical elasticity~\citep{Mura,Li} 
is recovered in Eqs.~(\ref{G}) and (\ref{A}). 

Using the Green tensor~(\ref{G}) and after a straightforward calculation
all the generalizations of the key-formulae~(\ref{B-grad-0})--(\ref{u-Burger-grad-0}) to gradient elasticity
are obtained.
The non-singular elastic distortion of a dislocation loop is given by
\begin{align}
\label{B-grad}
\beta_{ij}(\Bx)&=-\frac{b_k}{8\pi}\oint_L
\Big[\big(\epsilon_{jkl}\delta_{ir}-\epsilon_{rkl}\delta_{ij}
+\epsilon_{rij}\delta_{kl}\big)\pd_l \Delta
+\frac{1}{1-\nu} \,\epsilon_{rkl}\pd_l\pd_i\pd_j \Big] A(R)\, 
d L'_r\, .
\end{align}
This is the `Mura formula' for a dislocation loop in gradient elasticity.
Using the constitutive relation $\sigma_{ij}=C_{ijkl}\beta_{kl}$,  
we find the non-singular stress field produced by a dislocation loop
\begin{align}
\label{T-grad}
\sigma_{ij}(\Bx)&=-\frac{\mu b_k}{8\pi}\oint_L
\Big[\big(\epsilon_{jkl}\delta_{ir}
+\epsilon_{ikl}\delta_{jr}\big)\pd_l \Delta
+\frac{2}{1-\nu}\, \epsilon_{rkl}\big(\pd_i\pd_j-\delta_{ij}\Delta\big)\pd_l
\Big] A(R)\, d L'_r\, ,
\end{align}
which can be interpreted as the Peach-Koehler formula within the framework of 
gradient elasticity.
The key-formula for the non-singular displacement vector 
in gradient elasticity is obtained as
\begin{align}
\label{u-Burger-grad}
u_i(\Bx) = \frac{b_i}{8\pi}\, \int_S \Delta\pd_j A(R)\, d S'_j
+\frac{b_l\epsilon_{rlj}}{8\pi}\, \oint_L
\bigg\{\delta_{ij} \Delta -\frac{1}{1-\nu}\, \pd_i \pd_j   
\bigg\}\, A(R)\,  d L'_r\, ,
\end{align}                  
which is the Burgers formula in the framework of gradient elasticity of
Helmholtz type.                         
Eq.~(\ref{u-Burger-grad}) determines the displacement field of a single 
dislocation loop.
The Eqs.~(\ref{B-grad})--(\ref{u-Burger-grad}) are straightforward, simple and
closely resemble the singular solutions of classical elasticity theory.
In the limit $\ell\rightarrow 0$,  
the classical expressions~(\ref{B-grad-0})--(\ref{u-Burger-grad-0}) 
are recovered from Eqs.~(\ref{B-grad})--(\ref{u-Burger-grad}).
The expressions~(\ref{B-grad})--(\ref{u-Burger-grad}) retain most of the
analytic structure of the classical Mura, Peach-Koehler and Burgers formulae.
It is obvious that the expressions~(\ref{B-grad})--(\ref{u-Burger-grad}) are
given in terms of the elementary function $A(R)$ shown in Eq.~(\ref{A})
instead of the classical expression $R$.
The simplicity of our results is based on the use of gradient elasticity
theory of Helmholtz type.
Our results can be used in computer simulations of dislocation cores
at nano-scale
and in numerics of dislocation dynamics 
like fast numerical sums of the relevant fields
as used for the classical equations (e.g.~\citep{Sun,Li}).
They may be implemented in dislocation dynamics codes (finite element
implementation) and compared to atomistic models.

\section*{Acknowledgement}
The author gratefully acknowledges the grants of the 
Deutsche Forschungsgemeinschaft (Grant Nos. La1974/2-1, La1974/3-1).

\end{document}